\documentclass{neuthist18}

\def\Journal#1#2#3#4{{#1} {\bf #2}, #3 (#4)}

\def\EPJ{\em Eur. Phys. Jour.}

\def\NPB{{\em Nucl. Phys.} B}
\def\NPA{{\em Nucl. Phys.} A}
\def\PLB{{\em Phys. Lett.}  B}
\def\PRL{\em Phys. Rev. Lett.}
\def\PRD{{\em Phys. Rev.} D}
\def\PRC{{\em Phys. Rev.} C}
\def\MPL{\em Mod. Phys. Lett.}

\def\bb0{0 \nu \beta \beta}
\def\nmbnueb{\bar{\nu}_\mu \rightarrow \bar{\nu}_e}
\def\be{\begin{equation}}
\def\ee{\end{equation}}
\def\bea{\begin{eqnarray}}
\def\eea{\end{eqnarray}}

\newcommand{\Photo}{}

\begin{document}
\vspace*{4cm}
\title{Neutrino Mistakes:  Wrong tracks and Hints, Hopes and Failures}

\author{Maury C. Goodman }

\address{High Energy Physics Division, HEP362, Argonne, IL. 6039 USA}

\maketitle\abstracts{
In the last two decades, the field of neutrino physics has made enormous 
progress in measuring the strength and frequency of neutrino and 
antineutrino oscillations.  Along the way, there have been many instances 
of misunderstanding which led to wrong measurements or speculation for new 
features of neutrino physics that are not now accepted as correct.  This 
is part of the natural process of science, but given the well-accepted 
notion that we learn from our mistakes, it is worthwhile to look at some 
examples and see what the lessons might be.  With that goal in mind, I 
have a list of results which might be termed “neutrino 
mistakes,” with the 
fact in mind that there is no well-accepted definition of a mistake, and 
no unique threshold for counting something as a mistake when you change 
your mind after you obtain more information.  After making the list,
I 
chose seven of them to discuss.
No clear conclusions were drawn from this exercise, but some 
interesting issues regarding putative wrong results are discussed.
}

\section{Introduction}

In the last two decades, the field of neutrino physics has made enormous
progress in measuring the strength and frequency of neutrino and
antineutrino oscillations.  At the Neutrino History Conference,
this progress was described along with many other stories since before
the idea of the neutrino was formulated by Pauli in 1930.
In addition to this progress, there were some claimed discoveries
which did not hold up under scrutiny.  There were rumors of new
effects which generated interesting discussion.  And there were other
things which might fall under the category of {\it mistakes}.  The title,
``Neutrino Mistakes:  Wrong tracks and Hints, Hopes and Failures" was
given to me by the organizers, and is what this paper tries to 
describe.
 
\par A list of neutrino {\it mistakes}
considered for this paper is given in 
Table~\ref{tab:list}~\cite{bib:list}.
I prepared the list first, and 
then tried to come up with an algorithm for deciding what was chosen for
this particular list.  Wrong theories or theoretical frameworks
did not make 
it to the list.  The idea of neutrinos as hot dark matter in the
eV mass scale led to a number of experiments and could certainly have
fit the label of ``wrong tracks and hints, hopes and failures."  
Likewise, SU(5) led to a series of underground nucleon decay
experiments which certainly had a strong effect on the history
of neutrino oscillations.  Another class of possible mistakes
that were not included related to experimental hardware issues.
The catastrophic loss of phototubes at Super-Kamiokande was not
included, nor was the leaky  collapsed bag which led to the
demise of the IMB experiment, although interesting stories 
could be told about such episodes.  And finally I've restricted
this to the field of neutrinos, although a large number of 
seeming mistakes permeate our field, from direct dark matter
claims~\cite{bib:dama} and Cygnus X-3 
observations~\cite{bib:cygnus} to 
the split A2~\cite{bib:a2} and the Oops-Leon~\cite{bib:oops}.  
Besides instances of what I call a consequential semantic
issue, my
 a-posteriori definition for what was included on the list
would be
an experimental search for new $\nu$ physics for which there was an 
apparent 
error of the first kind or error of the second kind.
Such a mistake might be due to an unusual statistical fluctuation,
a systematic error that wasn't taken into account, a wrong interpretation
of good data, or a theoretical misunderstanding.  The threshold for
exploring new opportunities to find new physics is necessarily low.
As scientists, we are constantly asking ``What if ...?"  Some of
the alleged mistakes in the list are unpublished and
some were no more than rumors.  

%Table 01 
\begin{table}[hbt] 
\centering 
{\begin{tabular}{|lc|} 
\hline 
 SIN report of $\mu \rightarrow e \gamma$  &  \\ \hline 
High y anomaly  &  \\ \hline 
NuTeV helium bag events  &  \\ \hline 
Klapdor's $\bb0$ signal  &  \\ \hline 
LSND and eV ``sterile" neutrinos  &  \\ \hline 
IMB limit on $\nu$ oscillations  &  \\ \hline 
Alternating neutral currents  &  \\ \hline 
Reines-Sobel $\nu$ oscillations  &  \\ \hline 
Vannucci PS191 oscillations  &  \\ \hline 
BNL 776 $\&$ 816 oscillations  &  \\ \hline 
BEBC oscillations  &  \\ \hline 
HPWF ``super" events  &  \\ \hline 
Oscillations in Bugey  &  \\ \hline 
Majoron emission in $\bb0$ PNL/USC  &  \\ \hline 
SPT vs. V-A  &  \\ \hline 
Superluminal $\nu$s  &  \\ \hline 
17 keV $\nu$  &  \\ \hline 
NuTeV anomaly  &  \\ \hline 
Tritium endpoint negative $m^2$  &  \\ \hline 
Kolar events  &  \\ \hline 
Early atmospheric $\nu$ lack of polarization  &  \\ \hline 
MINOS $\bar{\nu} \theta_{23}$  &  \\ \hline 
God's mistake  &  \\ \hline 
$\nu$ grammar  &  \\ \hline 
Labels for $\Delta m^2_{ab}$  &  \\ \hline 
PDG m($\nu$) encoding  &  \\ \hline 
Which $\nu$ is a particle?  &  \\ \hline 
Karmen time anomaly  &  \\ \hline 
Time variation in Troitsk $m_\nu^2$  &  \\ \hline 
30 eV $\nu$ from ITEP  &  \\ \hline 
\end{tabular}} 
\caption{An unordered list of neutrino topics which might 
be regarded
as involving mistakes.}
\label{tab:list} 
\end{table}

Let me also list some of my personal guides in evaluating experimental
results:  1) there are an infinite number of tests of the null
hypothesis, 2) there is no theory of systematic error, 3) you can't
prove anything in physics, 4) the union of two confidence levels is
not a confidence level,  and 5) the commonly used 5$\sigma$ criterion
is based on misunderstandings and is wrongly used.

While the word ``mistake" carries negative connotations, for the
most part there should be no impugning of the scientists that
participated in the experiments mentioned in the list above.
We learn from our mistakes.  It is for that reason that this
subject was included in the 2018 Neutrino History Conference in Paris.  
But the
conclusion of this paper will present no unifying theme or lesson
from the topics considered here.  Instead, many of the cases described
raised a set of unique issues that may interest some readers.
But to repeat, it is not the goal here to criticize any physicist
or collaboration for reporting a result which is later
considered to be incorrect.

\section{Report of $\mu \rightarrow e \gamma $}

Since neutrinos have mass, there is a standard model
diagram for the unseen decay mode  $\mu \rightarrow e \gamma$.
The predicted branching fraction is:
\begin{equation}
B = 5 \times 10^{-48} [\Delta m^2_{21}(eV)^2]^2 \sin^2(\theta_{12})
\cos^2(\theta_{12})
\label{eq:mu}\end{equation}
We now know these neutrino mixing parameters so that this is of the order 
of $10^{-60}$.
When I was a graduate student in the 1970's, I heard a rather specific
rumor that this process had been seen by an experiment at the SIN facility
in Switzerland.  I never heard a talk about this result and in fact
when SIN published their results a few years later, they published
a limit, $B < 1.0 \times 10^{-9}$ at 90\% CL~\cite{bib:sin}.  In
that publication, they reported: ``No evidence for the existence
of the process has been found."  They also reported,
``The measured positron-photon energy distributions are completely
described by the decay $\mu^+ \rightarrow e^+ \nu_e {\bar{\nu}}_\mu 
\gamma$ and accidental coincidences."   I don't know if initial
interest in those events was the source of the rumor, but I mention
it as a possibility.
However a published indication of this rumor does exist in a
theoretical paper at that time on muon number nonconservation
by Bjorken and Weinberg~\cite{bib:bj}.  They wrote:
``{\it It would be disingenuous for us not to acknowledge that our
interest in this question was kindled by an experiment now in
progress at Schweizerisches Institut fur Nuklearfoschung [cf. Physics
Research in Switzerland, Catalog 1975 (Swiss Physical Society, Bern,
1975), p. 207], and by rumors of a positive signal.  However, our
considerations here do not depend on any assumptions about the eventual
outcome of this experiment; indeed, we believe that even if this
measurement were to yield a null result, it would be worthwhile to
push on to the greatest accuracy.}"  In fact, searches for this
process are continuing since any observation at a level larger than
that implied by Eq.~\ref{eq:mu} would indicate new physics in the
lepton sector.  
\par But at the time, this rumor led to a series of lectures at
Fermilab by Robert Shrock, who was then a postdoc in the Fermilab
theory group~\cite{bib:shrock}.  These lectures included a detailed
look at the time at the theoretical basis and phenomenology of
neutrino oscillations, and was where I learned about the subject
in detail for the first time.  For me personally, this rumored result,
which was wrong, never described and never published, was 
extremely useful in my career.

\section{Report of 17 keV neutrino}
\begin{figure}
\centerline{\includegraphics[width=0.7\linewidth]{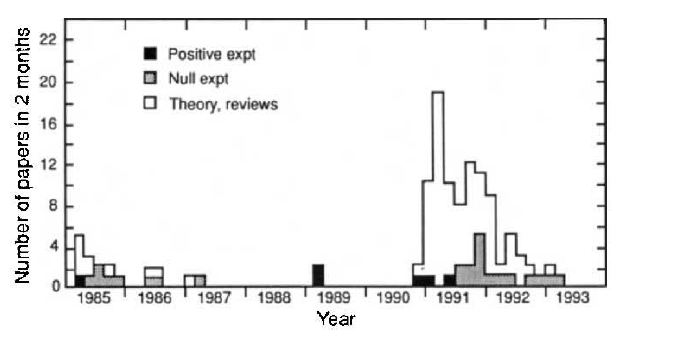}}
\caption[]{Timeline of positive and negative reports of
a 17 keV neutrino. \cite{bib:morrison}}
\label{fig:timeline}
\end{figure}

\begin{figure}
\centerline{\includegraphics[width=0.7\linewidth]{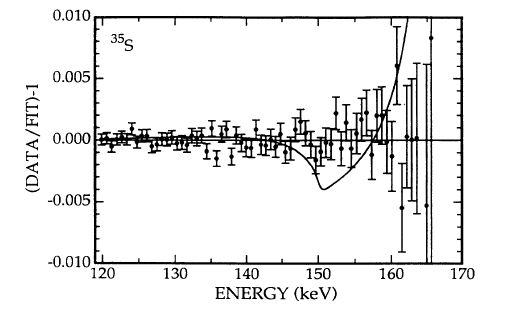}}
\caption[]{Shape function as reported by 
Mortara~\cite{bib:mortara}
in $^{35}$S with that expected if there was a 17 keV neutrino.}
\label{fig:stuart1}
\end{figure}

\begin{figure}
\centerline{\includegraphics[width=0.7\linewidth]{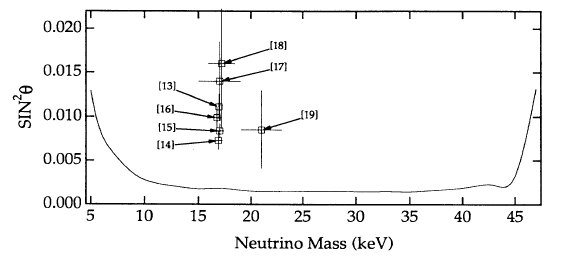}}
\caption[]{Limits on neutrino mass and mixing from 
Mortara\cite{bib:mortara}
along with values from previous positive reports.}
\label{fig:stuart2}
\end{figure}

Tritium decay is a popular object to use for studying the beta spectrum,
as the shape of the distribution near the 18.6 keV endpoint could be
sensitive to non-zero neutrino mass.  But in a study reported
in Simpson~\cite{bib:simpson1}, a kink was reported at 1.5 keV in the
spectrum, corresponding to a possible neutrino mass of 18.6 - 1.5 = 17.1 
keV with a mixing probability (P) of 3\%.  This was followed by
other experiments which failed to see a signal and set a limit at that
mass and P $<$ 0.3\%~\cite{bib:neg17}.  Then Hime and Simpson repeated the
search in $^{35}$S and reported a kink corresponding to 16.9 keV with
a mixing probability 0.7\% \cite{bib:simpson2}.  This was followed  with 
another 
experiment in $^{35}$S which reported an 8$\sigma$ observation at 17 keV
with P = 0.8\% \cite{bib:hime}.  A timeline of some positive and 
negative results is shown in Fig.~\ref{fig:timeline} \cite{bib:morrison}.
There were more  
nonobservations, but the definitive exclusion is considered as coming
from an experiment led by Stuart Freedman \cite{bib:mortara} in $^{35}$S
which reported the mixing probability of -0.0004 $\pm$ 0.0008(stat)
$\pm$ 0.0008(syst) consistent with zero.  Figure~\ref{fig:stuart1} shows
the residuals from a fit to the pileup corrected energy spectrum, along 
with the shape of an expected signal from a 17 keV neutrino.  
Figure~\ref{fig:stuart2} shows the 95\% upper limit on mixing as a 
function of neutrino mass along with the results from previous positive
experiments.   Both figures are from Mortara \cite{bib:mortara}.  
\par
\begin{figure}
\begin{minipage}{0.5\linewidth}
\centerline{\includegraphics[width=0.7\linewidth]{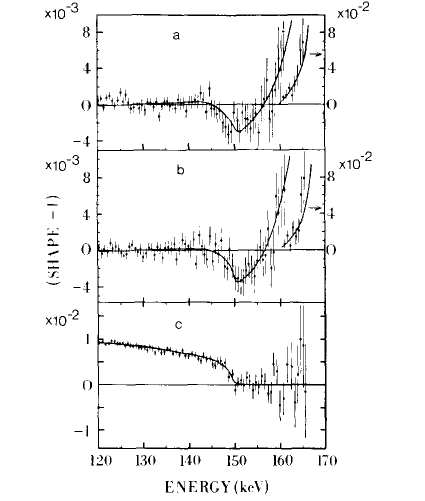}}
\end{minipage}
\hfill
\begin{minipage}{0.5\linewidth}
\centerline{\includegraphics[width=0.7\linewidth]{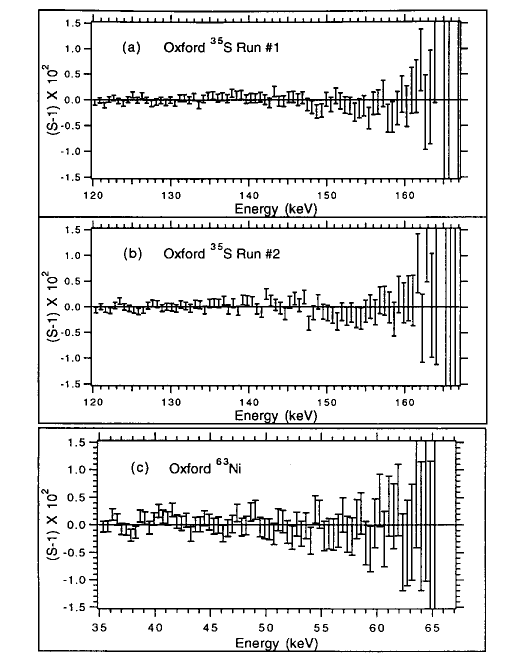}}
\end{minipage}
\hfill
\caption[]{Comparison of shape factors in $^{35}$S as first
reported in 
Hime and Jelley~\cite{bib:hime} and later in 
Hime~\cite{bib:hime2}}
\label{fig:hime}
\end{figure}

 Andrew Hime did further calculations which explained the
wrong signal in Hime and Jelley~\cite{bib:hime}.  He showed in 
Hime \cite{bib:hime2} that
scattering effects were likely responsible.  Compare the shape
factors in Figure~\ref{fig:hime} on the left from Hime and 
Jelley \cite{bib:hime}
with those on the right from Hime \cite{bib:hime2}.  The difference
involved a more complete electron response function with intermediate
scattering.

\section{Neutrinoless Double Beta Decay}
\par
\begin{figure}
\centerline{\includegraphics[width=0.7\linewidth]{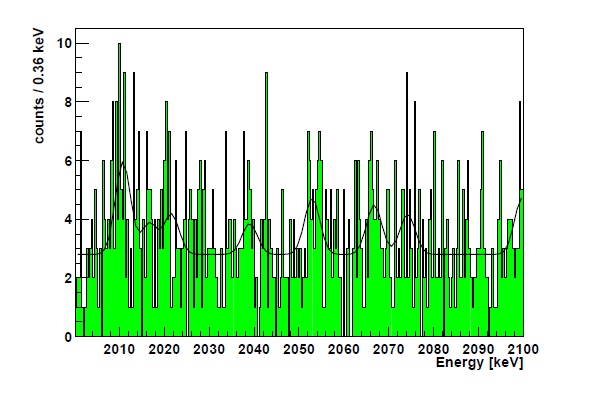}}
\caption[]{
Spectrum of the Heidelberg-Moscow experiment claim
for $\bb0$.
\cite{bib:hmyes2}}
\label{fig:hm}
\end{figure}

 In isotopes from a few select nuclei, decays can only happen by 
double beta
decay.  In the last 30 years, several measurements have been made of
two neutrino double beta decay.  But if the neutrino is a Majorana 
particle,
i.e. is its own antiparticle, then neutrinoless double beta decay
should also take place at rates that are low but predictable, given
other neutrino parameters, up to nuclear matrix elements.  There are
a variety of calculations of matrix elements which differ on a 
linear scale.  In 2001, the Heidelberg-Moscow collaboration, using
$^{76}Ge$ in an experiment at the Gran Sasso Lab, set a limit
on the lifetime for neutrinoless double beta decay
greater than 1.9 $\times$ 
$10^{25}$~y at 90\% CL \cite{bib:hmno}
in a paper signed by 14 authors.  Later that year, a subset of four 
authors claimed evidence for a signal with a lifetime 1.5 -0.7 +16.8 
$10^{25}$ y at 95\% CL \cite{bib:hmyes}.  This is shown in
Fig.~\ref{fig:hm} from Klapdor-Kleingrothaus~\cite{bib:hmyes2} where a fit 
finds an 
excess above background
at the known two electron energy, along with other known and
unknown lines.  
Soon thereafter, a critique of this claim appeared on the
arXiv with several authors from the nuclear beta decay community
and was 
later published \cite{bib:comment}.
They wrote, ``We discuss several limitations in the analysis provided
in that paper and conclude that there is no basis for the presented
claim."  
\par Mention of the evidence appeared in my January 2002
newsletter \cite{bib:lblc1} which goes to a large fraction of the neutrino
physics community.   I mentioned the lifetime, neutrino mass,
arXiv numbers from Heidelberg-Moscow's
positive and negative reports under the headline ``Evidence that
neutrinos are Majorana particles".  John Beacom wrote to me
\cite{bib:beacom} pointing out his own criticisms of the result
and opining that this report should not meet the standards of
my newsletter.  I replied to him that I didn't have standards but
I did have deadlines, a comment he has repeated back to me 
with a wry smile.  Then in my February newsletter, I reported
on the critique in the arXiv under the headline, ``Neutrino Mass may
not be .39 eV" \cite{bib:lblc2}.  I received an email from 
Klapdor-Kleingrothaus
\cite{bib:kk}
who was ``surprised to see that you handle the comment put on the
web as hep-ex/0202018 on the same level as our published paper..."
and ``May I propose that you better take out this unserious Comment
from your web page."  I responded that my newsletter gave equal attention 
to the discovery of neutrino oscillations and a novel about a Neanderthal 
neutrino physicist who appeared in the SNO detector.  Professor
Klapdor-Kleingrothaus never subscribed to the newsletter.
\par
\begin{figure}
\centerline{\includegraphics[width=0.7\linewidth]{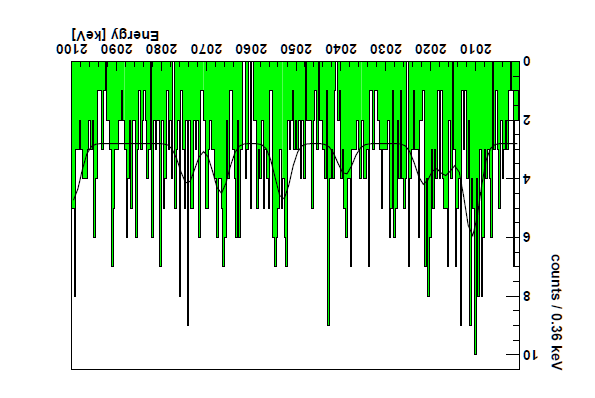}}
\caption[]{
Spectrum of the Heidelberg-Moscow experiment claim
for $\bb0$.
\cite{bib:hmyes2}}
\label{fig:hmusd}
\end{figure}

 This episode raised in my mind that there are a number of
unscientific factors which affect whether or not we believe a result
that we hear, particularly when we know that we don't have the background
or experience 
to
fully comprehend every scientific issue.   In this case, there were many
such factors.
The result was published before there was a preprint.
It was published in a journal on which one author was associated.
A significant fraction of collaboration didn't sign the paper.
One test of the validity of a {\it peak} in a distribution is to
look at the distribution upside-down and looking for a {\it dip}
(see Fig. \ref{fig:hmusd}). 
The only talk which
I heard from Klapdor-Kleingrothaus seemed arrogant
and he repeatedly touted this with the unrelated DAMA 
claimed discovery
of Dark Matter, another result that much of the community did and does
not believe.  There was a public argument with one of his
collaborators, and it was clear that he was not sharing the data
with them.
As we left that talk, 
Doug Michael,
a colleague of mine known for his insightful language, 
commented ``Even if it’s right, it’s wrong" \cite{bib:michael}.
I interpreted this to mean that he thought that if 
neutrinoless double beta decay exists in this channel, this
particular analysis had enough flaws to preclude being considered
a discovery.  My opinion is that the result is probably the
result of 
a-posteriori analysis.

\par
It didn't appear to me that this result was believed within the
neutrino community.
It did not lead theorists to write many papers despite their apparent 
predilection for
Majorana neutrinos.  But it couldn't be ignored either, and
the result became a benchmark for $\bb0$ experiments on their
way to achieving sensitivity to the non-degenerate inverted
neutrino hierarchy.  In the limits published by the  EXO-200
collaboration \cite{bib:bbexo}, GERDA collaboration \cite{bib:g}
and KamLAND-Zen collaboration \cite{bib:kz}, the limits
are specifically contrasted with this claimed positive result.
In the first two papers it is done in the conclusion, while
in the last it was done in the abstract.  This is entirely
appropriate, but I mention it as a point of irony. 

\section{Superluminal Neutrinos}
\par The OPERA report of neutrinos traveling from CERN to the
Gran Sasso with a velocity apparently faster than light led
virtually every particle physicist to have a story to tell.
I will start with a paper from the MINOS collaboration which
had the goal of measuring the time of flight of neutrinos
from Fermilab to the Soudan mine, 730 km in distance.  This was
potentially sensitive to a delay in arrival of low energy neutrinos
due to neutrino mass, though the sensitivity was far from
interesting on the mass scale relevant to atmospheric neutrino
oscillations.  
\par MINOS published a paper based on a study of the
arrival time of events at Soudan compared to the batch structure
of the beam for 473 events in the far detector \cite{bib:minosv}.
  The velocity
was measured to be (v-c)/c = 5.1 $\pm$ 2.9 10$^{-5}~eV^2$ at 68\%CL
which was used to limit the effective neutrino mass $m_\nu <
50~MeV/c^2$ at 99\% CL.   It is probably typical within the
field of High Energy Physics that we have not read a majority
of our own papers.  This is less true for neutrino
experimenters, as opposed to members of
collider collaborations, but this 
was
one paper on MINOS that I hadn't read when we published
it in 2007, and I certainly wasn't aware that we had an almost
two sigma superluminal result.  But in the course of repeating this
measurement with much greater accuracy, some members of the OPERA
collaboration were well aware of it.
\par In their first preprint, OPERA reported a measurement
 (v-c)/c = (2.48 $\pm$ 0.28 (stat.) $\pm$ 0.30 (sys.))$\times 10^{-5}$
in a preprint \cite{bib:opera1} 
dated 22 September 2011.  With
comparable statistical and estimated systematic errors, this was reported
as a 6$\sigma$ measurement.  On 23 September a seminar at CERN was
broadcast live on the web, at which it was reported that they had
obtained this result months before and tried to find an experimental
explanation before they presented it to the rest of the scientific
community.  On the same day, CERN issued a press release
\cite{bib:cernp} and interest in this result went way beyond
the scientific community.  There was some scrutiny of the result,
which led to a revised preprint on 17 November 2011 which was submitted
for publication, although the paper was never published
\cite{bib:opera2}.  A possible
loose connector was identified as the probable explanation on 25 February
2012, and the final study taking this into account appeared in the
arXiv on 12 July 2012 \cite{bib:opera4,bib:cernp}.

\par It seems to me that it was the press release from CERN,
and not the public seminar, which led to the world-wide attention for
this reported result.  As a form of {\it Gedanken history}, let
me imagine that the same preprint was released and seminar given
without the press release.  I would imagine that some science reporters
would have tracked down some scientists for comment, and that two or
three weeks later, there might have been an article in the science
section of the New York Times.  But the idea that Einstein might have
been wrong about the speed of light would not have reached everyone's
grandparents who have a TV or read the popular press.  The story
in the scientific community would have been the same.  The embarrassment
to science outside of the community would not have happened.

\par This point led to lively discussion at the History of Neutrino
meeting.  I have the unverifiable sense that a majority of our
colleagues wish that this result had not become a subject of the
popular and non-scientific press.  The point was brought up that it
becomes too easy for the public to disbelieve all science that they
want to disbelieve, when they have examples like this to use,
despite the careful caveats in the press release.  And this does
harm in places such the climate debate, where scientific and
non-scientific arguments get mixed up.  But the view was also
expressed that wide mention of our field does some good.
One
neutrino physicist mentioned that as a result of the reports of this 
experiment, his family expressed interest about what he was doing for the
first time.
\par The OPERA result led to a burst of interest and energy 
for MINOS to repeat its measurement 
with greater accuracy.  New techniques and
timing devices were deployed to reduce the systematic error and take
advantage of the greatly increased statistics in the MINOS far detector.
A new result was presented at a conference 
\cite{bib:minosv2} but a journal article 
took a long time to be published~\cite{bib:minosv3}.
This seems due to a combination of lack of interest from the referees
after the unpublished OPERA result was explained,
together with reduced time to deal with details 
 as competing demands  from elsewhere
took precedence.  

\section{LSND and eV sterile neutrinos}

\par
\begin{figure}
\centerline{\includegraphics[width=0.7\linewidth]{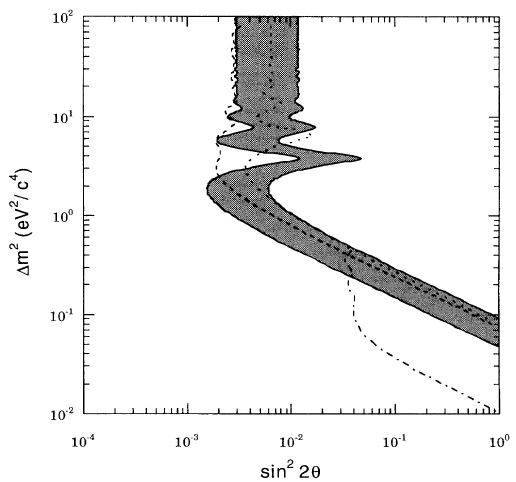}}
\caption[]{
Parameter space suggested by the original LSND publication
using a 2$\nu$ analysis.
\cite{bib:lsnd}}
\label{fig:lsnd}
\end{figure}

 
In 1994, the LSND experiment at Los Alamos presented an analysis
at the Neutrino 1994 conference showing an excess of 8 electron
neutrino candidates with a background of 0.9 consistent with
$\nmbnueb$ oscillations followed by inverse beta decay \cite{bib:lsnd1}.
The next year they published an excess of
16.4 +9.7 -8.9 $\pm$ 3.3 events \cite{bib:lsnd}.  The two neutrino
parameter space suggested by this excess is shown in Fig.~\ref{fig:lsnd}.
Not every member of the collaboration signed the paper.
In the same issue of Physical Review Letters, one author used the
same data employing cuts rather than a likelihood formula to obtain
an excess of 5 events with a background of 6.2, setting a limit
on $\nmbnueb$ \cite{bib:lsndhill}.  Additional data was reported in 
Aguilar \cite{bib:lsnda}.
While LSND was originally designed to search for non-zero values
of what we now call $\theta_{12}$, given the resolution of that
channel as related to the solar neutrino measurements, LSND is
now considered to be evidence for sterile neutrinos.  The MiniBooNE
experiment was designed to test this idea with different values
of L and $E_\nu$ but similar $L/E_\nu$.  Originally it reported
that its data was inconsistent with 
LSND~\cite{bib:miniboone1,bib:miniboonepress}.  
Further data with both neutrinos
and antineutrinos is interpreted by some as supporting the
LSND sterile neutrino idea \cite{bib:miniboone2,bib:lykken} but what 
appears
to have happened is that the data looks the same, while the
previous low-energy excess, outside the original blinded search
area, is now considered potential signal.  The case for sterile
neutrinos is sometimes buttressed with the 20\% Gallium anomaly
\cite{bib:gallium} and the 3\% reactor neutrino anomaly 
\cite{bib:reactor}.

\par Unlike most other results mentioned in this paper, there is
a sizable minority of neutrino physicists who consider that these results
motivate an aggressive continued search for sterile neutrinos as an
explanation of these short-baseline anomalies \cite{bib:sbp}.
It is not the role of this paper to evaluate the strengths and
weaknesses of the sterile neutrino interpretation of the anomalies.
A number of issues that would be part of such an evaluation would
be the apparent inconsistency of 0.3\%, 3\% and 20\% effects,
the Karmen result and Karmen's limit being better than its
sensitivity, the LSND decay in flight result, the
continued use of 2$\nu$ formulae, cosmological constraints,
inconsistency of $\nu_e$ appearance with $\nu_\mu$ disappearance
and limits from MINOS, NOvA and Ice-Cube.
\par
\begin{figure}
\centerline{\includegraphics[width=0.7\linewidth]{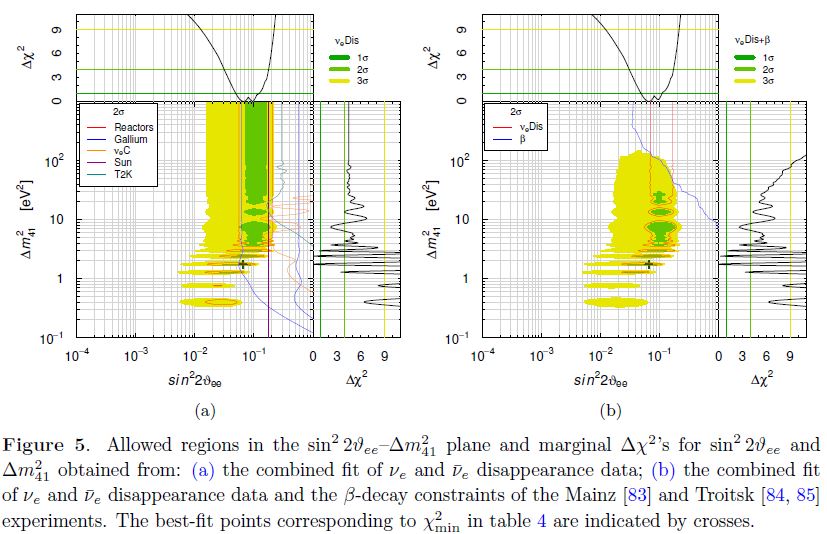}}
\caption[]{
Allowed region for sterile neutrino parameters from
a global fit in Gariazzo
\cite{bib:giunti}.}
\label{fig:giunti}
\end{figure}

 There are anomalies, which cannot be explained within the
3$\nu$ paradigm.  In Gariazzo~\cite{bib:giunti}, a global fit
to all the data is performed allowing for a sterile $\nu$ in the
3+1 scheme and a best fit is shown in Fig.~\ref{fig:giunti}.
But the best fit is a bad fit.  I have the impression that even
many of the advocates of sterile neutrino searches aren't
confident that the answer lies in 3+1 sterile neutrinos, but
rather that something more complicated is going on that might
involve new physics, and that sterile neutrinos might play part
of the answer.  But that makes a definitive experiment impossible.
If you don't know what you are looking for, you might find it,
and might not find it, but you cannot logically rule it out.
With this in mind, I coined an answer to the simple question
why physicists disagree:  {\it If the data doesn't agree with
the null hypothesis or the alternative hypothesis, some say you need 
more data, while some say you need more hypotheses.}

\section{IMB neutrino oscillation limit}
\par
\begin{figure}
\centerline{\includegraphics[width=0.7\linewidth]{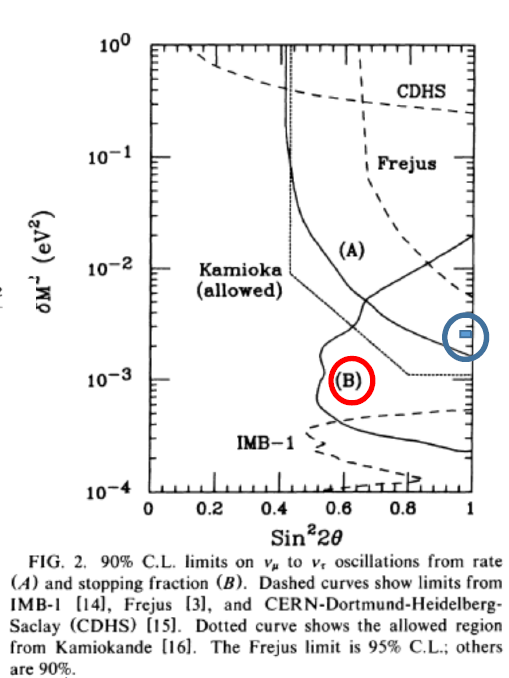}}
\caption[]{
Region ruled out by analysis in Becker-Szendy~\cite{bib:imb1}
(B) with the current allowed region added in blue.}
\label{fig:imb}
\end{figure}

 In 1992, IMB published a neutrino oscillation limit based on the 
ratio of upward-going stopping $\mu$ from atmospheric $\nu$ to upward 
going $\mu$ \cite{bib:imb1}.  Based on this analysis, they ruled out a 
region of
parameter space labeled B in Fig.~\ref{fig:imb}.  As shown by the blue
rectangle which I have added to the figure from 
~Becker-Szendy \cite{bib:imb1},
that is just the region in which neutrino oscillations turned out to be.
As it became clear this limit must be wrong, an explanation for why
it was misleading was sought.  Part of the IMB collaboration submitted
an abstract to the 1999 International Cosmic Ray Conference 
\cite{bib:imb2} suggesting that a wrong neutrino cross section model
was used.  But the contents of and results from ``a more realistic
cross section model" did not appear in the proceedings.  
One part
of this story is that long-baseline neutrino experiments
were being proposed and compared in the mid 1990's and the IMB
limit was used to argue for shorter baselines than were in fact
needed.  The argument was unsuccessful and the longest-baseline
choice was chosen (i.e. MINOS over BNL's P889).  Due to matter
effects and resolution of the mass ordering, it turns out that
even longer baselines would have been desirable, and that is
part of the motivation for the DUNE program.

\section{God's mistake}
The search for neutrino oscillations has benefited from
fortuitous experimental sensitivity to parameters, which led
to a light-hearted argument circa 1995 that there was an
intelligent design of neutrino parameters \cite{bib:idesign}.
The argument is that god made: the optimum choice for
$\Delta m^2_{21}$ = 8.2 $\times~10^{-5}~eV^2$ which gives
a resonant MSW effect transition in the middle of the solar
neutrino energy spectrum;  an optimum choice for $\theta_{12}
\sim~30^\circ$ causing the effect to be big enough to be seen
in KamLAND;  an optimum choice for $\Delta m^2_{32}$ = 2.3 
$\times~10^{-3}~eV^2$ giving the transition from no oscillation
to full oscillation in the middle of the range of possible
distances that atmospheric neutrinos travel to get to the
detector; the optimum choice for $\theta_{23}~\sim~\pi/4$
leading to the dramatic large effects that were easy to see in
atmospheric neutrinos; and the then unknown
$\theta_{13}$ being small enough not to confuse the interpretation
of the above, so that the story up to that time could be 
adequately explained
with 2$\nu$ formulae.  But the acid test was mentioned as whether
$\theta_{13}$ would be large enough to see CP violation and determine
the mass ordering.  This was confirmed in 2011-2012 when the
reactor neutrino experiments Double Chooz, Daya Bay and RENO found
$\theta_{13}$ to be as large as possible consistent with the previous
limits.
\par So in 2012 I extrapolated the intelligent design concept to the
still unanswered questions about neutrinos.  This implied (1) the
CP violation parameter $\delta~\sim~3\pi/2$ to most quickly determine
the mass ordering and to get large CP violation; (2) the inverted
mass order so that we can more readily measure $\bb0$ to distinguish
Dirac and Majorana neutrinos, and perhaps measure the beta decay
endpoint, and (3) neutrinos should be Majorana which seems to be
the more interesting case for theorists, and we want our theorists
to be happy.
\par Question 3 hasn't been answered yet, but early comparisons
of T2K, NOvA and reactor data suggest $\delta~\sim~3\pi/2$ may be
close to the answer.  However there is increasing evidence that the
mass order is normal, in contradiction to the apparent
``Intelligent 
Design"
answer.  Did god make a mistake?  The more likely answer is that
the normal mass order is just what we want and we aren't
intelligent enough to realize why yet.

\section{Discussion of Issues raised}
In this paper I have 
examined a 
number of historical contexts and issues regarding some wrongly
interpreted results.  However, 
there is no firm conclusion about how to recognize a wrong result
or how to proceed after one is presented to the community.  Just
as serendipity often leads to a breakthrough (searches for nucleon
decay leading to the discovery of neutrino oscillations is an
obvious example), so too some of the examples I gave had positive
consequences.  But there are clear downsides, and
time spent pursuing 
mistaken results that could
have been more usefully spent elsewhere is impossible to estimate.

\par A number of questions have come up in this exercise.
Is it fair that a wrong result can become a benchmark and get a huge
number of citations?  When should a possible paradigm shift become
actively publicized outside our community?  What should referees do
with a result they don't believe is scientifically accurate but don't
know what mistake was made, if any?  When should a collaboration stop
trying
to resolve a new hard-to-believe result internally and announce it?
Are rumors useful or counterproductive?  How should we view papers
which a full collaboration does not sign?  Should there be more
active skeptics of hard-to-believe results, or is the fate of being
ignored satisfactory?

\par Some of the examples in Table 1 are published, some are presented at 
conferences, and some
only make it to the rumor stage.  Some elicit prompt critics,
and some are faced with a ``let's wait and see" attitude, till
they are refuted or ignored.  Some are followed by published 
retractions
or explanations, and others are only followed by a loss
of interest once a 
definitive
exclusion becomes well known.  Some motivate creative theoretical
speculations while others fail to motivate any such ideas.  Some
lead to numerous follow-up experiments and others fail to do so.
Some lead to useful thinking outside the box, and others do not.
Could the understanding of the rightness or wrongness of
these examples have been made more quickly?

\par There is one mistake in my view which is common in our
field and that relates to the so-called 5$\sigma$ criterion for
discovery.  We often consider a null hypothesis which is that the
data can be understood without new physics, and a particular new
effect as the alternative hypothesis.  We design a test statistic
that is sensitive to the difference and quota a chance probability
that the data is described by the null hypothesis, usually turning
P into x$\sigma$, assuming a Gaussian probability distribution.
But this is only valid if the hypothesis and statistic are specified
a-priori.  Of course we do a-posteriori analysis all the time.  It is
part of our job to look for unusual aspects of the data.  But while
the number of $\sigma$ is calculated in an identical way for an
a-priori and a-posteriori hypothesis, the meaning is totally different.
I cringe when I hear colleagues dismiss an a-priori 3$\sigma$ effect and
demand 5$\sigma$ because ``I've seen so many 3$\sigma$ effects go away."
Those were likely all calculated a-posteriori.
\par I suspect that as our interesting paradigms change, there is a strong
time-dependent effect on our mistakes.  Fifty years ago, a curious
neutrino result might have been interpreted as SPT deviations from
the V-A theory \cite{bib:shrock2}.  Today the same result might be
analyzed as a sterile neutrino.
\section{Conclusion}

While calling a result a mistake has a connotation of 
criticism, I do not 
criticize the vast majority of these reported 
results.  While we want to avoid noise, sharing results we don't
understand sooner rather than later might help the field get to 
the truth in a more efficient way.  Our field of particle physics does a 
poor job in my opinion of
presenting statistical arguments in a consistent way.
In particular it is often difficult for an outsider to distinguish 
between a x$\sigma$ effect calculated from an 
a-priori test and an x$\sigma$ effect calculated from an a-posteriori 
test.
We also do a poor job explaining to ourselves and others how we 
conclude anything based on 
whatever combination of data, theory and instinct that we use.\
Nevertheless,
we seem to do an excellent collective job of taking seriously results 
which get vindicated and being skeptical of results which do not.
That is probably the best test of how well our field is doing.
And once again, the field of neutrino physics has been thriving.

%\subsection{Photograph}

%You may want to include a photograph of yourself below the title
%of your talk. A scanned photo can be 
%directly included using the default command\\
%\verb^\newcommand{\Photo}{\includegraphics[height=35mm]{mypicture}}^\\
%just before the 
%\verb^\begin{document}^
%line. If you don't want to include your photo, just comment this line
%by adding the \verb^%^ sign at the beginning of 
%the line and uncomment the next one
%\verb^%\newcommand{\Photo}{}^ by removing its \verb^%^ sign.

%\begin{figure}
%\begin{minipage}{0.33\linewidth}
%\centerline{\includegraphics[width=0.7\linewidth,draft=true]{figexamp}}
%\end{minipage}
%\hfill
%\begin{minipage}{0.32\linewidth}
%\centerline{\includegraphics[width=0.7\linewidth]{figexamp}}
%\end{minipage}
%\hfill
%\begin{minipage}{0.32\linewidth}
%\centerline{\includegraphics[angle=-45,width=0.7\linewidth]{figexamp}}
%\end{minipage}
%\caption[]{same figure with draft option (left), normal (center) and 
%rotated (right)}
%\label{fig:radish}
%\end{figure}

\section*{Acknowledgments}

I would like to acknowledge helpful comments from
Evgeny Akhmedov, Zelimir Djurcic, John LoSecco, Naba Mondal,
Jurgen Reichenbacher, Jack Schneps, Phil Schreiner, Robert Shrock, Hank 
Sobel, Daniel Vignaud and Cosmas Zachos.  But all mistakes in this
discussion of mistakes are my own.

\section{Other}
\par I will try to briefly describe some of the other ``mistakes" 
from Table~\ref{tab:list} not already 
covered more fully.  I divide
them into 3 categories:  (1) other reports which could be 
interpreted as oscillations which didn't pan out; (2) a few
issues which may be regarded as semantic but I feel have some 
substance,
and (3) everything else.
\subsection{Other oscillation reports}
A reactor experiment by Reines reported a strange CC/NC ratio and showed 
an
allowed parameter space for neutrino oscillations \cite{bib:rsobel}.
Hank Sobel thinks that a changed cross section might have been responsible
\cite{bib:rsobel2}.  An early ITEP measurement of the tritium spectrum
was consistent with a 30 eV mass neutrino \cite{bib:itep30}.  Later 
measurements seemed to give negative values for $m^2$ \cite{bib:negm2}.
And a CERN beam dump result was interpreted as $\nu_e \rightarrow 
\nu_\tau$
in De Rujula~\cite{bib:cerndump}.

\subsection{Substantive semantic issues}
For many years, the PDG reported neutrino masses associated
with flavors, such as electron or tau.  We now know that's like
saying ``the hole that the electron went through in the two slit
experiment".  This was fixed in 2003 \cite{bib:shrockpdg}.
A related issue is that $\nu_e, \nu_\mu and \nu_\tau$ are flavor
eigenstates, but not particles, which are the mass eigenstates.
The PDG updated its chart of fundamental particles at th beginning of
the 21st century \cite{bib:pdgc}, but CERN and Fermilab, among others,
haven't fixed this in their graphics.  Many people still confuse
$\Delta m^2_{23}$ and $\Delta m^2_{32}$.  These differ by a sign and
the difference is one of the main goals of new experiments.  It is 
possible to define them with the opposite convention than usually done
(in the usual convention $\Delta m^2_{21}$ is positive) but I don't think
this is the actual source of the mistake.

When long-baseline or short-baseline is used as an adjective, which
is almost always, there needs to be a hyphen.

\subsection{Other neutrino mistakes}
 Unexpected y distributions (y = $E_{had}/E_\nu$) 
were 
reported in FNAL E1 at kinematic low x \cite{bib:e1}.  This
was contradicted by CCFR and Charm \cite{bib:nohighy}.  The Kolar 
experimented reported events deep underground consistent with decays
of a new particle in the air outside the detector \cite{bib:kolar}.
This was never confirmed.  A search at NuTeV for supersymmetric 
particles decaying in a Helium bag found three events~\cite{bib:hebag}
which did not match the signal hypothesis.  The 
original atmospheric neutrino flux 
calculations used for the “ratio of ratios” did not take into account the 
fact that the muon is polarized \cite{bib:atmop}.  Even after this was 
corrected, this was cited as a reason not to take the atmospheric
neutrino anomaly seriously.    An oscillation signal was reported by Bugey 
2
at Neutrino 1984 with $\Delta m^2~=~0.2~eV^2$ \cite{bib:nu84}.  Two 
contradictory (different L/E) positive results from BNL 
involvied low energy electron excesses in a $\nu_\mu$ 
beam. 
BNL 776 saw 23 $\nu_e$ compared to 13.1 expected and
BNL 816 saw 110 $\nu_e$ compared to 53 expected.  Both were
reported at Neutrino 1988~\cite{bib:nu88}.  When Experiment 1 first saw 
three $\mu$ events,
there were seminars about ``super" events, but no claim was ever
published \cite{bib:cline}.  There was a timing anomaly in Karmen,
which was interpreted as a massive new particle \cite{bib:karment},
but this was ruled out by Daum et al. \cite{bib:daum} and shown
by J. Reichenbacher to be caused by neutrons \cite{bib:juergen}.
A double beta decay experiment reported evidence for a Majoron
\cite{bib:majoron1}, which was contradicted in \cite{bib:majoron2}
and the argument continued in \cite{bib:majoron3}.  It does not 
seem to be a topic of current interest.  When MINOS measured neutrino
oscillation properties separately for $\nu$ and $\bar{\nu}$, the 
numbers looked different for $\theta_{23}$ \cite{bib:minos1}.  
MINOS invented an a-posteriori test and quoted the difference as 2\% 
chance probability, or about 2.4 $\sigma$.  MINOS then asked for and
received additional $\bar{\nu}$ running and the discrepancy disappeared
\cite{bib:minos2}.  Before neutral currents were firmly established,
there were conflicting results known as ``alternating neutral currents"
\cite{bib:nc1,bib:nc2,bib:nc3,bib:nc4}.  More recently, the 
NuTeV collaboration measured the NC/CC ratio
to determine the Weinberg angle and got an unexpected result 
\cite{bib:nutev} known as the NuTeV anomaly.


\begin{thebibliography}{99}
\bibitem{bib:list}  Many items in this list are referenced when discussed
in this paper.  Others are from my memory and are not discussed further.
\end{thebibliography}
\end{document}